\documentclass[prl,aps,twocolumn,superscriptaddress,showpacs]{revtex4}
\usepackage{graphicx}
\begin{document}

\title{Low energy electronic states and triplet pairing in layered cobaltates}

\author {     G. Khaliullin }
\affiliation{ Max-Planck-Institut f\"ur Festk\"orperforschung,
              Heisenbergstrasse 1, D-70569 Stuttgart, Germany }
\affiliation{ E. K. Zavoisky Physical-Technical Institute of the
              Russian Academy of Sciences, 420029 Kazan, Russia } 
\author {     W. Koshibae }
\affiliation{ Institute for Materials Research, Tohoku University, 
              Sendai 980-8577, Japan }
\author {     S. Maekawa }
\affiliation{ Institute for Materials Research, Tohoku University, 
              Sendai 980-8577, Japan }

\date{\today}

\begin{abstract}
The structure of the low-energy electronic states in layered cobaltates 
is considered starting from the Mott insulating limit. 
We argue that the coherent part of the wave-functions 
and the Fermi-surface topology at low doping 
are strongly influenced by spin-orbit coupling of the correlated 
electrons on the $t_{2g}$ level. An effective $t$-$J$ model 
based on mixed spin-orbital states is radically different 
from that for the cuprates, and supports unconventional, 
pseudospin-triplet pairing. 
\end{abstract}


\maketitle

The layered cobalt oxides Na$_x$CoO$_2$  
exhibit a number of remarkable properties with potential applications.  
At large $x$, they are unconventional metals showing 
a large thermopower (suppressed by magnetic field \cite{Wan03})   
and also signatures of localized magnetic states. 
Such a mixture of itinerant transport and Curie-Weiss 
magnetism (``Curie-Weiss metal" \cite{Foo03}) 
is not easy to reconcile, and there is growing evidence that special 
charge/orbital ordering correlations are at work at large $x$ \cite{Ber04}. 
Static charge ordering is observed at $x=0.5$ \cite{Foo03}. 
These observations are not totally unexpected, 
as a rich interplay between spin/charge/orbital degrees of freedom 
is a common phenomenon in transition metal oxides. Yet a real surprise 
and new challenge for theory was the recent discovery \cite{Tak03} of 
superconductivity at $T_c\sim 5$~K in water-intercalated cobaltates 
with low sodium content about $x\sim 0.3$. It is believed that the 
superconductivity of the cobaltates emerges from a correlated metallic state 
with enhanced electronic mass \cite{Cho03,Uem04}, and it may have 
an unconventional, possibly spin-triplet \cite{Wak03,Fuj04,Hig03}   
pairing symmetry.    
  
Theoretically, regarding Na$_x$CoO$_2$ as spin 1/2 Mott insulator 
doped by spinless charge carriers, a $t$-$J$ model similar to that for 
the high-$T_c$ cuprates has been considered \cite{Bas03,Kum03,Oga03,Wan04}. 
However, the predicted time reversal violating  
$d_1+id_2$ singlet state is not supported by experiment \cite{Uem04,Hig03}, 
and other proposals employing fermi-surface nesting and/or charge-density 
fluctuations have been discussed \cite{Tan03,Tan04} in favor of  
spin-triplet pairing.     
       
In this Letter, we show that the relevant $t$-$J$ model 
for the cobaltates is in fact 
qualitatively different from its simplest version used 
in Refs.\cite{Bas03,Kum03,Oga03,Wan04}. The key point is that the low-energy
electronic states of the CoO$_2$ layer are derived from the Kramers pseudospin 
doublets of the Co$^{4+}$ ion 
with {\it mixed spin and orbital} quantum numbers. 
The projection of the (initially spin-antiferromagnetic) 
$J$-interaction on the pseudospin states has nontrivial consequences 
for the internal structure of the Cooper-pairs, leading 
to novel {\it pseudospin-triplet} pairing.  

The basic structural element of Na$_x$CoO$_2$ are CoO$_2$ layers, which 
consist of edge sharing CoO$_6$ octahedra slightly compressed along 
the trigonal axis \cite{Lyn03}. 
The Co ions form a 2D triangular lattice, sandwiched by oxygen layers. 
Sodium doping is believed to introduce spinless Co$^{3+}$ states 
into the spin 1/2 Co$^{4+}$ background.    

A minimal model for the cobaltates should include the orbital degeneracy 
of the Co$^{4+}$-ion \cite{Kos03}, where a hole in 
the $d^5(t_{2g})$-shell has the freedom to occupy one out of three  
orbitals $a=d_{yz}$, $b=d_{xz}$, $c=d_{xy}$. The degeneracy is partially
lifted by trigonal distortion, which stabilizes the $A_{1g}$ electronic 
state $(a+b+c)/\sqrt{3}$ over the $E'_g$-doublet 
$(e^{\pm i\phi}a+e^{\mp i\phi}b+c)/\sqrt{3}$ (hereafter $\phi=2\pi/3$): 
$H_{\Delta}=\Delta [n(E'_g)-2n(A_{1g})]/3$. 
The value of $\Delta$ is not known; Ref.\cite{Kos03} estimates it 
$\sim$25~meV, while the band structure calculations for a related 
structure give $\sim$100~meV \cite{Ezh98}. 

In terms of the effective angular momentum $l_{eff}=1$ of $t_{2g}$-shell  
\cite{Abr70}, the functions $A_{1g}$ and $E'_g$ correspond to 
the $|l_z=0\rangle$ and $|l_z=\pm 1\rangle$ states, respectively. 
Therefore, a hole residing on the $E'_g$ orbital doublet will experience an  
unquenched spin-orbit interaction $H_{\lambda}=-\lambda(\vec{l}\cdot\vec{s})$. 
The constant $\lambda$ for Co$^{4+}$ is about 
640 cm$^{-1}\approx 80$meV \cite{Bla83}. Although $\lambda$ 
is somewhat smaller than the bare hopping matrix element 
$t\sim 0.1$~eV in cobaltates (inferred from the free-electron bandwidth 
$\sim$1 eV \cite{Sin00}), the spin-orbit coupling  
strongly affects the coherent motion of the fermions  
when the quasiparticle bandwidth is reduced 
by correlation effects to much smaller values  
$\sim 0.1$~eV \cite{Has03}.   

\begin{figure} 
\includegraphics[clip,width=7.2cm]{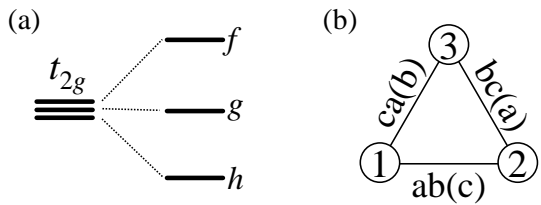} 
\caption{
${\bf (a)}$ The $t_{2g}$-orbital degeneracy of Co$^{4+}(d^5)$-ion is 
lifted by trigonal distortion and spin-orbit interaction.  
A hole with pseudospin one-half resides on the $f$-level. 
Its wavefunction contains both $E'_g$ and $A_{1g}$ states 
mixed up by spin-orbit coupling. 
${\bf (b)}$ The hopping geometry on the triangular lattice of Co-ions. 
$\alpha \beta (\gamma)$ on bonds should read as $t_{\alpha \beta}=t$, 
$t_{\gamma \gamma}=-t'$, and $\alpha,\beta,\gamma \in \{a,b,c\}$ 
with $a=d_{yz}$, $b=d_{xz}$, $c=d_{xy}$. The  
$t$-hopping stems from the charge-transfer process via oxygen ions,  
while $t'$ stands for the direct $d$-$d$ overlap.    
}
\label{fig1}
\end{figure}

At $x=0$, $H=H_{\Delta}+H_{\lambda}$ is diagonalized by a transformation:
\begin{eqnarray}
\label{eq1}
\alpha_{\sigma}=i[c_{\theta}e^{-i\sigma \psi_{\alpha}}f_{-\bar\sigma}+ 
is_{\theta}f_{\bar\sigma}+e^{i\sigma \psi_{\alpha}}g_{\bar\sigma}+ 
\nonumber \\ 
s_{\theta}e^{-i\sigma \psi_{\alpha}}h_{-\bar\sigma}-
ic_{\theta}h_{\bar\sigma}]/\sqrt{3}~, 
\end{eqnarray}
where $c_{\theta}=\cos\theta, s_{\theta}=\sin\theta$, 
$\alpha=(a,b,c)$ and $\psi_{\alpha}=(\phi, -\phi, 0)$,
correspondingly. The angle $\theta$ is determined from  
$\tan{2\theta}=2\sqrt{2}\lambda/(\lambda + 2\Delta)$. As a result, one 
obtains three levels, 
$f_{\bar\sigma}, g_{\bar\sigma}, h_{\bar\sigma}$ [see Fig.\ref{fig1}(a)],  
each of them are Kramers doublets with pseudospin one-half $\bar\sigma$. 
The highest, $f$-level, which accommodates a hole in the $t_{2g}$-shell, 
is separated from the $g$-level by   
$\varepsilon_f-\varepsilon_g=
\lambda+\frac{1}{2}(\lambda/2+\Delta)(1/\cos{2\theta}-1)$. 
This splitting is $\sim 3\lambda/2$ at $\lambda \gg \Delta$, and 
$\sim\lambda$ in the opposite limit. It is important to observe that 
the pseudospin $f_{\bar\sigma}$ states 
\begin{eqnarray}
\label{eq2}
|\bar\uparrow \rangle_f&=&ic_{\theta}|+1\rangle|\downarrow\rangle - 
s_{\theta}|0\rangle|\uparrow\rangle, \nonumber \\
|\bar\downarrow \rangle_f&=&ic_{\theta}|-1\rangle|\uparrow\rangle - 
s_{\theta}|0\rangle|\downarrow\rangle 
\end{eqnarray}
are coherent mixtures of different orbital and spin states, 
and this will have important consequences for the symmetry 
of the intersite interactions, as we see below.       

We model the band motion by the following Hamiltonian suggested 
by the edge-shared structure \cite{Kos03}: 
\begin{equation}
\label{hopping}
H_t^{ij}=t(\alpha^{\dagger}_{i\sigma}\beta_{j\sigma}+
\beta^{\dagger}_{i\sigma}\alpha_{j\sigma})-
t'\gamma^{\dagger}_{i\sigma}\gamma_{j\sigma} + h.c.~, 
\end{equation}
where $t=t_{pd}^2/\Delta_{pd}$ originates from the 
$d$-$p$-$d$ process via the charge-transfer gap $\Delta_{pd}$, 
and $t'>0$ is the direct $d$-$d$ hopping.  
On each bond, there are two orbitals ($\alpha,\beta$)  
active in $d$-$p$-$d$ process, while the third one ($\gamma$) 
is transfered via the direct $d$-$d$ channel [Fig.\ref{fig1}(b)].
The hopping geometry in real situation could, of course, be   
more complicated (e.g., include $p$-$p$ hoppings). The $t,t'$-model 
is the simplest possibility chosen for illustrative purposes. 

The quasiparticle band structure is obtained within a slave boson approach, 
where the electron operator is represented as $\alpha_{i\sigma}\Rightarrow 
e^{\dagger}_i \bar\alpha_{i\sigma}\Rightarrow  
\sqrt{x}\bar\alpha_{i\sigma}$+{\it incoherent part}. 
The last equation implies that we consider a fermi-liquid regime, 
where holons (described by the $e^{\dagger}_i$ operator) are condensed, 
while the $\bar\alpha_{i\sigma}$ operators represent coherent 
fermionic quasiparticles. On the mean-field 
level (the incoherent part being discarded), this 
leads to $t\rightarrow t_{coh}\simeq xt$, and one obtains  
a renormalized hopping Hamiltonian $H_t^{coh}\simeq xH_t$.  
While such a Gutzwiller-type approximation fails to capture high-energy 
electronic processes far from the Fermi level, it is believed to provide 
a qualitatively correct description of the low-energy states. 

\begin{figure}
\includegraphics[clip,width=8.8cm]{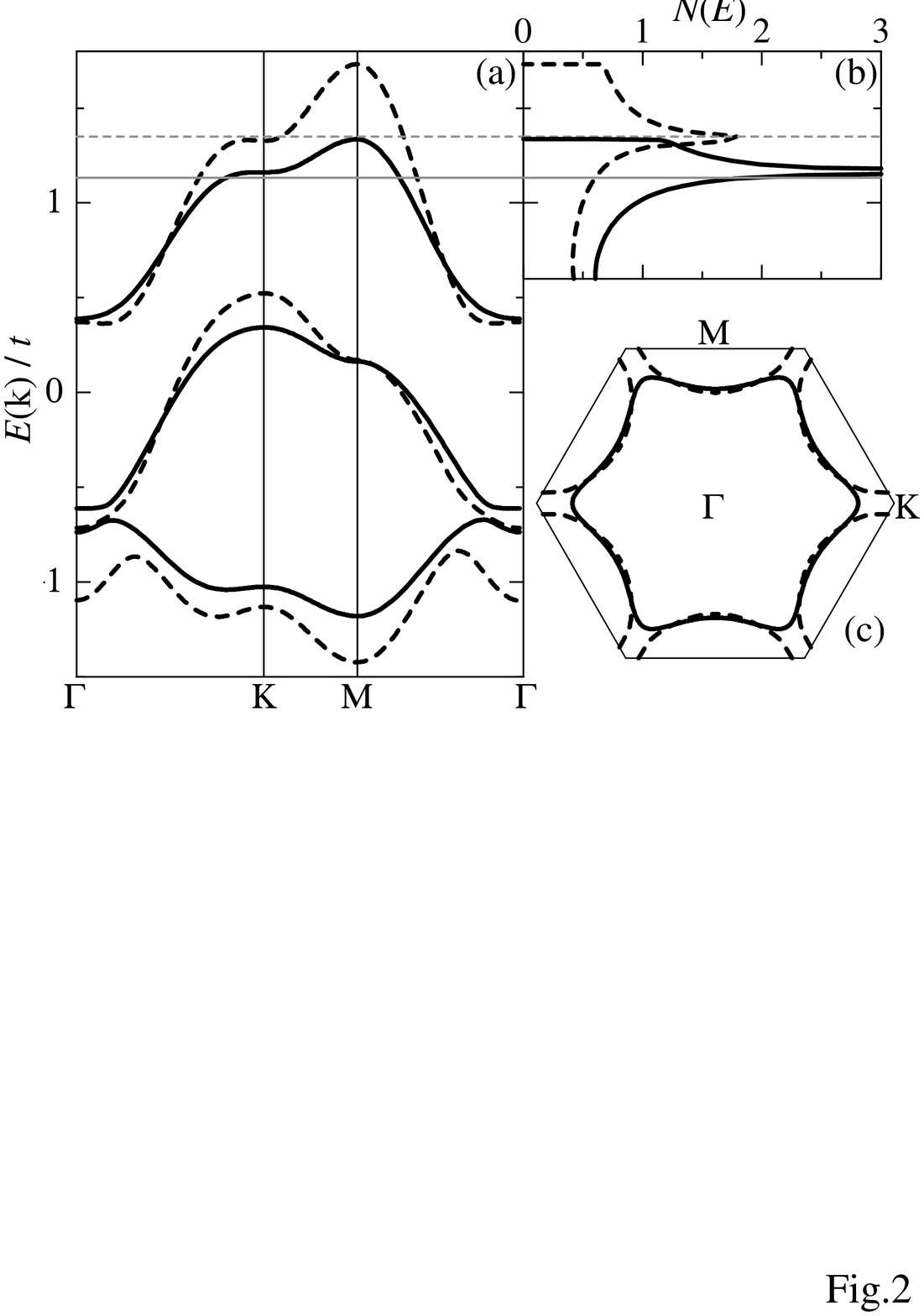}
\caption{${\bf (a)}$ Quasiparticle dispersion curves 
in a slave-boson mean-field approximation.  
${\bf (b)}$ Density of states (
in units of $1/t$) near the 
Fermi level. The latter is indicated by thin gray lines.  
${\bf (c)}$ Fermi surface in the Brillouin zone. 
$\Gamma=(0,0)$, $M=(0,2\pi/\sqrt 3)$, $K=(4\pi/3,0)$. 
In all the panels, solid (dashed) lines  
correspond to the doping level $x=0.2$ ($x=0.3$). 
Parameters used: $t'/t=0.8$, $\lambda/t=0.7$, $\Delta/t=1.0$.   
}
\label{fig2}
\end{figure}

Under the transformation (\ref{eq1}), the hopping Hamiltonian obtains 
a rather complicated matrix structure in the full spin/orbital space, and 
the elements of this matrix sensitively depend on the angle $\theta$. 
The Hamiltonian $H_{\Delta}+H_{\lambda}+H_t^{coh}$ has been diagonalized 
in momentum space numerically.  
The obtained quasiparticle dispersion curves are shown in Fig.\ref{fig2}(a). 
The first point to notice is that even for 
a substantial doping level, $x=0.3$, there is a clear separation of bands 
which can be traced back to the on-site level structure discussed above. 
In particular, we find at small doping that the states near the 
Fermi level are derived dominantly from the $f$-pseudospin states; therefore, 
they are dispersive modes with mixed spin-orbital quantum numbers. 
The "$f$"-band has the bottom at the $\Gamma$-point \cite{note1},     
and shows flat portions near the $K$-points 
in the Brillouin zone of the triangular lattice, which lead to a singular
density of quasiparticle states near the Fermi level [Fig.\ref{fig2}(a,b)]. 
An interesting doping evolution of the Fermi surface (FS) [Fig.\ref{fig2}(c)]
is also related to the presence of those flat portions.   
The overall energy window, covered by three quasiparticle bands, is reduced 
to $\sim 3t$ by correlation effects. The high energy, incoherent 
electronic states, extending up to a free-electron scale $\sim 9t$,   
are beyond the present approximation. The consideration of the 
incoherent part of the electronic motion, which is a difficult task in 
correlated models in general, is further complicated here because of the  
additional orbital degrees of freedom. 

As far as the low-energy physics at small doping is concerned, 
we arrived in fact at a single-band picture for states 
near the FS. 
However, this band has little to do 
with a conventional single-orbital band of $A_{1g}$ symmetry 
suggested by a free-electron band calculations \cite{Sin00} and 
taken in Refs.\cite{Bas03,Kum03,Oga03,Wan04} as a starting point 
to develop $t$-$J$ model physics for cobaltates. 
The crucial difference here is that the quasiparticle wave-functions 
are made of states which do not conserve 
neither spin nor orbital angular momentum separately; rather, they 
are build on eigenstates of the total angular momentum, as reflected 
in Eq.(\ref{eq2}). This unusual situation results from the interplay 
between strong spin-orbit coupling of Co$^{4+}$ ion and 
the quasiparticle kinetic energy reduced by correlation effects.    
The main message is that spin-orbit coupling becomes  
increasingly effective near the Mott insulating limit and is thus 
essential for the formation of the quasiparticle bands, and hence for  
the FS 
topology. This implies that the FS 
shape at low doping may {\it not necessarily be similar} to that 
given by a free-electron picture, as one would 
expect in single-band systems like the cuprates. In a more general context, 
the outlined picture might be relevant also to other transition metal 
oxides, where narrow quasiparticle bands are derived  
from (quasi)degenerate $t_{2g}$-states with strong 
spin-orbit coupling, such as compounds based on late-3$d$, 4$d$ and 5$d$-ions. 
  
Now we turn to the superexchange interactions which, in analogy 
with high-$T_c$ cuprates, could be one of the relevant interactions 
responsible for superconducting pairing.        
When the FS 
states are derived from pseudospin 
states 
with mixed spin and orbital quantum numbers, an important question  
arises about 
implications of such a mixing for the pairing symmetry. 
Projected on the pseudospin states, 
the superexchange interactions 
may in fact give nontrivial pairing channels, which were not present 
in the original, pure-spin Heisenberg model.  

A superexchange Hamiltonian in orbitally degenerate 
systems reads in general as 
$H_J=J[(\vec S_i\cdot\vec S_j)\hat A_{ij}+\hat K_{ij}]$ \cite{Kug82}.  
The energy scale is $J=4t^2/E$, where the virtual charge excitation 
energy $E$ is determined either by the on-site Coulomb $U_d$ or 
the charge-transfer energy $\Delta_{pd}$, depending on which one is lower.   
The operators $\hat A_{ij}, \hat K_{ij}$ represent the orbital degrees of
freedom. Without orbital degeneracy (e.g., in the cuprates) 
$\hat A_{ij}=1, \hat K_{ij}=-1/4$, and $H_J$ 
supports uniquely singlet pairing. In the present situation, 
one has to: ({\it i}) derive a full structure of the orbital operators 
$\hat A_{ij}$ and $\hat K_{ij}$ (which usually depends on 
the $\langle ij \rangle$-bond orientation in crystal via the 
hopping geometry \cite{Kug82}), 
and ({\it ii}) project the obtained $H_J$ onto the active 
pseudospin $f_{\bar\sigma}$-subspace given by Eq.(\ref{eq2}). Details of this 
lengthy derivation \cite{Kha04} will be presented elsewhere; for our purpose, 
the result can conveniently be represented in the following form:  
\begin{equation}
\label{HJf}
H^f_J(ij)=-J_f \kappa_s s_{ij}^{\dagger}s_{ij}
-J_f \kappa_t T_{ij}^{\dagger} T_{ij}. 
\end{equation}
Here, $J_f=J/9$ with $J=4t^2/\Delta_{pd}$. In equation (\ref{HJf}), 
the interactions are separated into pseudospin singlet and triplet 
channels, namely, 
$s_{ij}=(f_{i\bar\uparrow}f_{j\bar\downarrow}-
f_{i\bar\downarrow}f_{j\bar\uparrow})/\sqrt{2}$, 
while 
\begin{equation}
\label{Tij}
T_{ij}=a_z t_{ij,0} + 
a_{xy}(e^{i\phi_{\delta}} t_{ij,1}+ 
e^{-i\phi_{\delta}} t_{ij,-1})/\sqrt 2~, 
\end{equation}
with $t_{ij,0}=i(f_{i\bar\uparrow}f_{j\bar\downarrow}+
f_{i\bar\downarrow}f_{j\bar\uparrow})/\sqrt{2}$~, 
$t_{ij,1}=f_{i\bar\uparrow}f_{j\bar\uparrow}$ and 
$t_{ij,-1}=f_{i\bar\downarrow}f_{j\bar\downarrow}$. 
The phase $\phi_{\delta}$ in Eq.(\ref{Tij}) and below depends on 
the $\langle ij \rangle$-bond direction $\delta$:  
$\phi_{\delta}=(0,\phi,-\phi)$ on $\delta=(12,23,13)$-bonds (see Fig.1b), 
respectively. The relative weights of the different components 
($M=0,\pm$1 projections of the total pseudospin of the pair) 
are controlled by the angle $\theta$, defined 
above, via $a_z=(1+\cos{2\theta})/2$ and $a_{xy}=\sin{2\theta}$. 
Finally, a constant  
\begin{eqnarray}
\label{kappa}
\kappa_s=\frac{1}{2}\left(\frac{3\cos{2\theta}-1}{2}\right)^2+
\frac{\Delta_{pd}}{U_d}
\left(\frac{3\cos{2\theta}-1}{2}+\frac{t'}{t}\right)^2\!, 
\end{eqnarray}
while for the triplet channel one has $\kappa_t=3/2$.
 
In general, the interaction (\ref{HJf}) supports both  
singlet and triplet pairings on the fermi-surface. 
Certainly, there are also some other pairing forces in the cobaltates; 
yet it is interesting to explore the outcome of the superexchange 
interactions alone. We have therefore calculated the mean-field 
superconducting transition temperatures as function of the parameters 
involved in our effective $t$-$J$ model for the $"f"$-pseudofermion band. 

In the singlet channel, it is known \cite{Bas03,Kum03,Oga03,Wan04} 
that pairing on the triangular lattice is optimized by a complex 
order parameter 
$\langle s_{ij} \rangle =D~e^{i\phi_{\delta}}$ (the degenerate conjugate 
state is obtained by $\phi_{\delta} \rightarrow -\phi_{\delta}$). 
In momentum space, the gap function is determined by 
$\gamma_d(\vec{k})=\gamma_1(\vec k)+i\gamma_2(\vec k)$, 
where 
$\gamma_1
=\cos k_x -\cos\frac{k_x}{2}\cos\frac{\sqrt{3}k_y}{2}$ and  
$\gamma_2
=\sqrt{3}\sin\frac{k_x}{2}\sin\frac{\sqrt{3}k_y}{2}$. 
This state breaks the time-reversal symmetry. 

As suggested by the very form of Eq.(\ref{Tij}), the bond 
dependence of the triplet order parameter is best parametrized 
according to the projection $M$ of the Cooper-pair pseudospin. Namely, 
a positive interference among the different $M$-channels is achieved by   
\begin{equation}        
\langle t_{ij,0}\rangle =d_z~, ~~~~~
\langle t_{ij,\pm1}\rangle =d_{xy} e^{\mp i\phi_{\delta}}.  
\end{equation} 
The pairing amplitudes $d_z, d_{xy}$ are proportional to $a_z$ and $a_{xy}$, 
respectively. The bond arrangement of the phases $e^{-iM\phi_{\delta}}$ 
translates in momentum space into the form-factors $\gamma_{z}(\vec{k})$ and  
$\gamma_{xy}^{\pm}(\vec{k})=\pm \gamma_x(\vec{k})+i\gamma_y(\vec{k})$, 
where 
\begin{eqnarray}
\gamma_z(\vec{k})&=&\sin k_x-2\sin\frac{k_x}{2}\cos\frac{\sqrt{3}k_y}{2}~, \\
\gamma_x(\vec{k})&=&\sqrt{3}\cos\frac{k_x}{2}\sin\frac{\sqrt{3}k_y}{2}~, \\
\gamma_y(\vec{k})&=&\sin k_x+\sin\frac{k_x}{2}\cos\frac{\sqrt{3}k_y}{2}~. 
\end{eqnarray}
The gap function has no nodes (apart from the $\Gamma$- and $M$-points), 
and the superconducting gap anisotropy is given by 
$\Delta(\vec k)\propto 
\sqrt{\left|a_z \gamma_z\right|^2+\left|a_{xy} \gamma_{xy}^{\pm}\right|^2}$ 
which depends on angle $\theta$.  
Remarkably, the pseudospin-triplet state is nondegenerate and 
thus {\it respects} the time-reversal symmetry. 
This is because the orbital currents associated with 
$M=+1$ and $M=-1$ components flow in opposite directions 
(observe $\pm$ signs in above equations),  
and the orbital angular momentum of the Cooper-pair is quenched.  
   
The mean-field $T_c$ for the triplet state  
(assuming the holons are already condensed) is obtained from 
\begin{equation}
1=J_f\sum_{\vec{k}}\frac
{\left|a_z\gamma_z\right|^2+\left|a_{xy}\gamma_{xy}^{\pm}\right|^2} 
{2\xi_{\vec k}}\tanh\frac{\xi_{\vec k}}{2T_c}~, 
\end{equation}
where $\xi_{\vec k}$ is a "f"-quasiparticle energy.  
To obtain $T_c$ in the singlet  
channel, one simply has to replace the numerator in this equation by  
($\frac{2}{3}\kappa_s |\gamma_d(\vec k)|^2$). 

\begin{figure}
\includegraphics[clip,width=8.8cm]{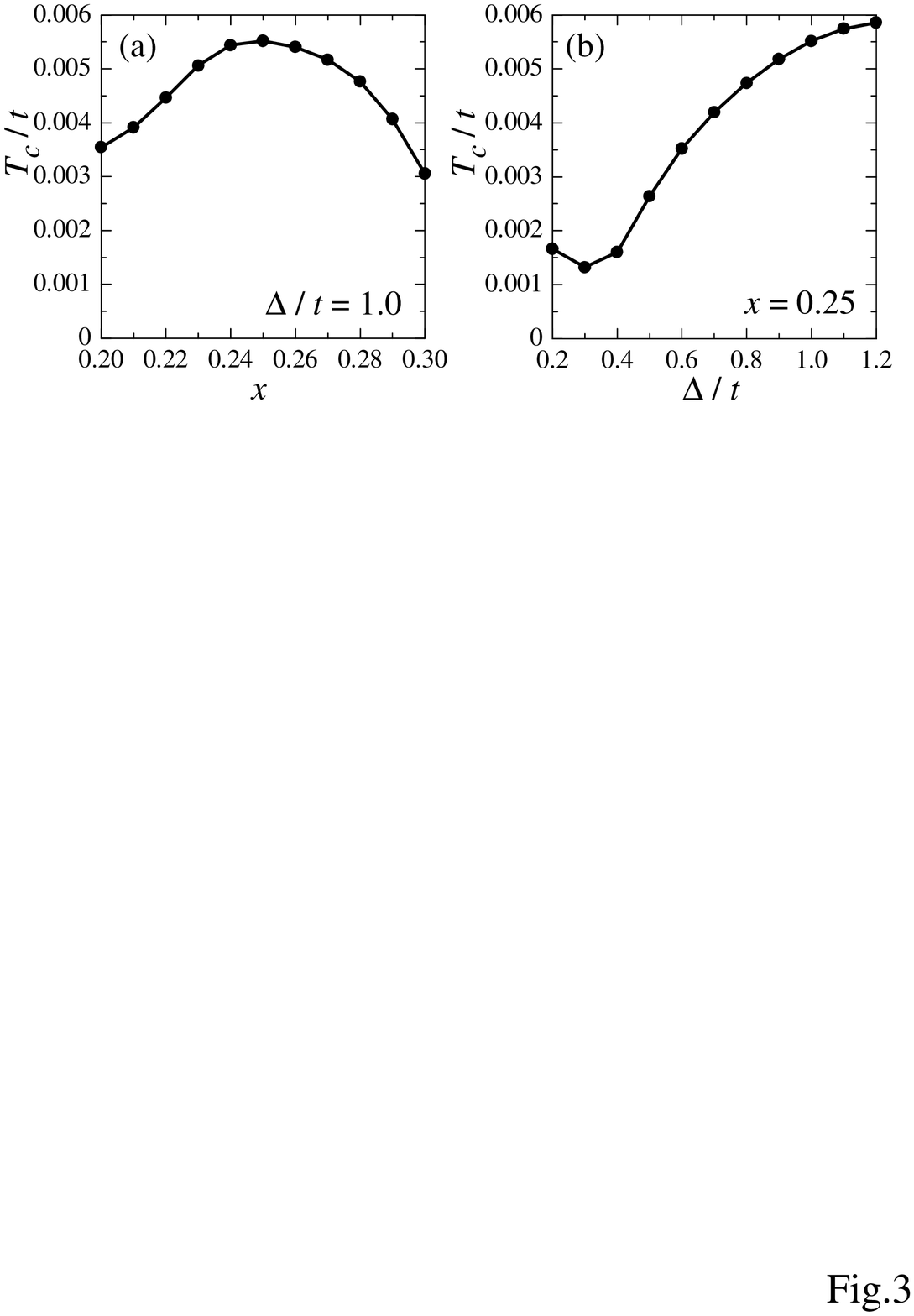}
\caption{${\bf (a)}$ Doping $x$ and ${\bf (b)}$ trigonal field splitting 
$\Delta$ dependences of the mean-field superconducting transition 
temperature $T_c$. Parameters used: $t'/t=0.8$, $\lambda/t=0.7$, $J_f/t=0.05$. 
}
\label{fig3}
\end{figure}

Numerical calculations show that for $t'<t$ and $\Delta_{pd}/U_d<1$ 
(relevant for cobalt oxides \cite{Zaa85}) the triplet state is 
always favored. Shown in Fig.\ref{fig3} are some plots for $T_c$, calculated 
for an effective exchange parameter $J_f=0.05t$ (implying $\sim$5 meV for 
$t\sim 0.1$~eV). Nonmonotonic dependences of $T_c$ on 
doping and 
the orbital splitting (induced by $c$-axis compression) are due to  
sensitive variations of the quasiparticle band near the $K$-points 
of Brillouin zone (Fig.\ref{fig2}). We note that the obtained $T_c$-values 
($\sim 5$~K for $t\sim 0.1$~eV) and their dependences are 
qualitatively consistent with experimental data \cite{Tak03,Lyn03}. 

The superexchange driven triplet state is interesting on its own. 
This state is stabilized by spin-orbit coupling acting on the 
low-energy states of doped Mott insulator with $t_{2g}$ orbital degeneracy.  
Its basic features (no time-reversal symmetry breaking, 
only partial Khight-shift drop below $T_c$, high critical 
fields $H_{c2}$, {\it etc}) are reminicent to reported data in 
cobaltates. However, a specific comparison with (ongoing and still 
controversial) experiments would require that some other  
relevant physics (electron-phonon coupling, charge-ordering 
tendencies that are enhanced by orbital-polaron effects \cite{Kil99}, 
phase separation, {\it etc}), may need to be included in a more 
realistic theory. We believe that the present work should be a proper
starting point for such a theory.    
   
To conclude, the low-energy Fermi-surface states in the underdoped cobaltates 
are strongly influenced by spin-orbit interaction 
intrinsic to the $t_{2g}$ electrons of 
Co$^{4+}$ ions. This leads to nontrivial symmetry of the superexchange
interactions and to a novel pseudospin-triplet paired state, 
qualitatively different from that in the high-$T_c$ cuprates. 
Rather, there might be some common physics between cobaltates 
and ruthenates \cite{Mac03} and also the recently discovered 
superconductors KOs$_2$O$_6$ \cite{Yon03}, 
where Ru(4$d$)- and Os(5$d$)-electrons with strong spin-orbit 
coupling reside on nearly degenerate $t_{2g}$ levels. 
 
The authors are grateful to B. Keimer, T. Tohyama and 
C. Honerkamp for useful discussions. 
One of us (G.Kh.) would like to thank the International Frontier Center 
for Advanced Materials at Tohoku University for its kind hospitality. 
This work was supported by Priority-Areas Grants from the Ministry 
of Education, Science, Culture and Sport of Japan, NAREGI and CREST.



\begin{thebibliography}{}

\bibitem{Wan03} Y. Wang, N.S. Rogado, R.J. Cava, and N.P. Ong, 
Nature {\bf 423}, 425 (2003).

\bibitem{Foo03} M.L. Foo {\it et al.}, 
\prl {\bf 92}, 247001 (2004).

\bibitem{Ber04} C. Bernhard {\it et al.}, 
\prl {\bf 93}, 167003 (2004).

\bibitem{Tak03} K. Takada {\it et al.}, 
Nature {\bf 422}, 53 (2003).

\bibitem{Cho03} F.C. Chou {\it et al.}, 
\prl {\bf 92}, 157004 (2004).

\bibitem{Uem04} Y.J. Uemura {\it et al.}, cond-mat/0403031. 

\bibitem{Wak03} T. Waki {\it et al.}, cond-mat/0306036. 

\bibitem{Fuj04} T. Fujimoto {\it et al.}, 
\prl {\bf 92}, 047004 (2004). 

\bibitem{Hig03} W. Higemoto {\it et al.}, cond-mat/0310324.

\bibitem{Bas03} G. Baskaran, \prl {\bf 91}, 097003 (2003).

\bibitem{Kum03} B. Kumar and B.S. Shastry, 
Phys. Rev. B {\bf 68}, 104508 (2003).

\bibitem{Oga03} M. Ogata, J. Phys. Soc. Jpn. {\bf 72}, 1839 (2003).  

\bibitem{Wan04} Q.-H. Wang, D.-H. Lee, and P.A. Lee, 
Phys. Rev. B {\bf 69}, 092504 (2004).

\bibitem{Tan03} A.Tanaka and X. Hu, 
\prl {\bf 91}, 257006 (2003).

\bibitem{Tan04}
Y. Tanaka, Y. Yanase, and M. Ogata, J. Phys. Soc. Jpn. {\bf 73}, 319 (2004);  
O.I. Motrunich and P.A. Lee, Phys. Rev. B {\bf 69}, 214516 (2004).

\bibitem{Lyn03} J.W. Lynn {\it et al.}, 
Phys. Rev. B {\bf 68}, 214516 (2003).  

\bibitem{Kos03} W. Koshibae and S. Maekawa, 
\prl {\bf 91}, 257003 (2003).

\bibitem{Ezh98} S.Yu. Ezhov {\it et al.}, 
Europhys. Lett. {\bf 44}, 491 (1998).

\bibitem{Abr70} A. Abragam and B. Bleaney, {\it Electronic Paramagnetic 
Resonance of Transition Ions} (Oxford University Press, New York, 1970).

\bibitem{Bla83} K.W. Blazey and K.A. M\"uller, 
J. Phys. C: Solid State Phys. {\bf 16}, 5491 (1983). 

\bibitem{Sin00} D.J. Singh, Phys. Rev. B {\bf 61}, 13397 (2000).

\bibitem{Has03} M.Z. Hasan {\it et al.}, 
\prl {\bf 92}, 246402 (2004); 
H.-B. Yang {\it et al.}, \prl {\bf 92}, 246403 (2004).

\bibitem{note1} If we project the hopping Hamiltonian (\ref{hopping}) 
onto the $A_{1g}$ states alone 
(consider $\alpha_i\rightarrow A_{1g,i}/\sqrt 3$), 
the result would be just opposite: $\Gamma$-point on the top.   

\bibitem{Kug82} K.I. Kugel and D.I. Khomskii, 
                Sov. Phys. Usp. {\bf 25}, 231 (1982). 

\bibitem{Kha04} G. Khaliullin (to be published). 

\bibitem{Zaa85} J. Zaanen, G.A. Sawatzky, and J.W. Allen, 
                 \prl {\bf 55}, 418 (1985).

\bibitem{Kil99} R. Kilian and G. Khaliullin, 
                  \prb {\bf 60}, 13458 (1999).

\bibitem{Mac03} K.K. Ng and M. Sigrist, 
Europhys. Lett. {\bf 49}, 473 (2000); 
A.P. Mackenzie and Y. Maeno, 
Rev. Mod. Phys. {\bf 75}, 657 (2003).

\bibitem{Yon03} S. Yonezawa, Y. Muraoka, Y. Matsushima, and 
Z. Hiroi, J. Phys.: Condens. Matter {\bf 16}, L9 (2004).  

\end{thebibliography}
\end{document}